# Universal Quantum Interconnects via Phase-Coherent Four-Wave Mixing

*Hao Zhang*[*1, 2], *Yang Xu*[3], *Linshan Sun*[1], *Wei Cui*[6], *Robert W. Boyd*[3, 5, 6] *and Sergio Carbajo*[*1, 2, 4]

[1]*Department of Electrical and Computer Engineering, UCLA, 420 Westwood, Los Angeles, CA 90095, USA*
[2]*SLAC National Accelerator Laboratory, 2575 Sand Hill Rd, Menlo Park, CA 94025, USA*
[3]*Department of Physics and Astronomy, University of Rochester, 500 Wilson Blvd, Rochester, New York 14627, USA*
[4]*Physics and Astronomy Department, UCLA, 475 Portola Plaza, Los Angeles, CA 90095, USA*
[5]*The Institute of Optics, University of Rochester, 275 Hutchison Rd, Rochester, New York 14627, USA*
[6]*Department of Physics, University of Ottawa, 150 Louis-Pasteur Pvt, Ottawa, Ontario K1N 6N5, Canada*
*haozh@g.ucla.edu

**ABSTRACT** Quantum transduction, which enables the coherent conversion of quantum information between disparate physical platforms, is a cornerstone for realizing scalable and interoperable quantum networks. Among various approaches, parametric frequency mixing processes such as four-wave mixing (FWM) offer a promising pathway toward efficient and low-noise transduction. In this work, we demonstrate the feasibility of coherent quantum state transfer by indirectly verifying high-fidelity wavefunction's phase mapping (>99%) from the input field to the generated output field wave. Using a gas-filled hollow-core capillary fiber, we systematically investigate spectral phase evolution across a broad range, including infrared (IR) to ultraviolet (UV) transitions, as well as conversions from telecom-band (1550 nm) to visible (516 nm) and deep-UV (308 nm) wavelengths. Our results reveal that strong phase coherence can be maintained throughout these diverse conversion regimes. Because quantum properties such as coherence and entanglement are intrinsically encoded in both the amplitude and phase of a photonic wavefunction, preserving spectral phase is essential for faithful quantum information transfer. We further show that efficient and phase-preserving transduction can be achieved by tuning system parameters, offering valuable insights into nonlinear coupling dynamics. These findings establish a promising foundation for advancing FWM-based quantum transduction schemes and open new avenues for integrating heterogeneous quantum systems across wide spectral domains within future quantum communication networks.

## 1. Introduction

The rapid advancement of quantum technologies hinges on the ability to efficiently interface and interconnect a wide variety of quantum systems [1,2]. Quantum transduction plays a pivotal role in enabling future quantum architectures by bridging disparate physical platforms [3,4], thereby supporting distributed quantum computing [5,6], hybrid quantum devices[7], and advanced quantum sensing[8]. A wide range of quantum transduction methods has been developed to meet the demands of quantum networking[4], where transducers serve as critical interfaces between quantum processors and optical communication channels. For example, the conversion of quantum information between microwave and optical domains[9–13], is essential for linking superconducting quantum processors with long-distance optical communication channels. Conventional approaches, such as piezo-optomechanical and electromechanical transducers, utilize mechanical modes as intermediaries to facilitate this conversion. While recent experiments[10] have demonstrated conversion efficiencies on the order of $(0.88 \pm 0.16) \times 10^{-5}$ with signal-to-noise ratios (SNR) exceeding unity, theoretical models suggest the potential for much higher efficiencies, exceeding 50%, and in some cases approaching near-unity levels, especially when employing impedance-matched designs or advanced materials such as antiferromagnetic topological insulators. Despite ongoing challenges, including mechanical noise, cavity losses, and the stringent requirement to preserve quantum coherence, recent progress underscores the central importance of transduction technologies in integrating heterogeneous quantum platforms within future architectures[7,14,15].

Frequency upconversion techniques have demonstrated both high efficiency and excellent coherence in converting infrared photons to the visible spectrum [16,17]. Materials such as silicon carbide and quantum dots have emerged as promising candidates for bridging classical and quantum photonics, owing to their favorable optical properties and compatibility with scalable device fabrication [17–20]. These advances in quantum transduction are propelling the development of quantum communication networks, optical quantum computing platforms, telecom interconnects, and distributed quantum architectures [21–24], bringing the vision of large-scale, interconnected quantum systems closer to reality.

Key insights from these studies can be summarized as follows: coherence preservation is often inferred indirectly through proxies such as conversion efficiency and added noise, with some approaches predicting near-unity efficiency under ideal conditions of intraband entanglement[25]. Direct measurements of input fidelity remain limited, reported explicitly in only a few cases, notably the acoustic bus system and teleportation-based protocols[26]. Scalability also varies significantly across platforms: electro-optic systems, while compact, encounter difficulties in maintaining high nonlinearity and quality factors at scale, whereas acoustic, phononic, and spin-based architectures offer greater potential for scaling to large qubit numbers and enabling reconfigurable quantum connectivity[27–29]. These studies highlight that no single method currently optimizes all desired metrics simultaneously, and performance comparisons are further complicated by the differing figures of merit employed across platforms.

Amid the various approaches to quantum transduction, including optical-to-microwave, optomechanical, and electro-optic schemes, photonic quantum transduction, particularly in the form of direct optical-to-optical frequency conversion, has emerged as a promising and technically scalable solution as part of the quantum transduction ecosystem. All-photonic transduction enables coherent conversion between photons of vastly different energies while preserving essential quantum attributes such as phase coherence and entanglement. This capability is critical for bridging heterogeneous quantum systems, such as linking atomic quantum memories with fiber-based telecom networks[4,17,30–32]. Of particular relevance, this process enables the combination of atomic quantum memories with fiber optic communication channels[33]. In such networks, the optimal transmission wavelengths in fibers, which lie in the infrared spectrum of telecommunications wavelengths, may significantly differ from the wavelengths at which quantum information can be locally generated, stored, or processed—typically in the UV to blue spectrum range[34–36]. The substantial energy difference (~2.4 eV) between the UV and telecom wavelengths has constrained state-of-the-art quantum converters to quasi-phase matching in periodically poled lithium niobate (PPLN) waveguides [34,37]. Still, the performance and error tolerance suffer substantially for practical long-haul quantum network connectors due to their high susceptibility to photorefractive damage.

Alternative nonlinear processes are being explored to overcome these material and performance limitations, with stimulated emission such as four-wave mixing (FWM) transduction emerging as a particularly promising candidate[30,38–45]. FWM has demonstrated measurable advantages in both theoretical and experimental studies[46]. For example, in a five-level atomic system, conversion efficiencies have reached approximately 33% under specific conditions and approach unity with increasing input power, accompanied by a reduction in the required conversion length[47]. Fiber- and intracavity-based models further reveal that FWM techniques can achieve lower noise levels and more accurate quantum state transfer compared to squeezed phase fluctuations, which in some cases lack explicit noise quantification or suffer from excess noise due to parametric amplification[48]. Furthermore, the spatiotemporal quasi-phase matching (QPM) in broadband fiber-based FWM has also enabled reconfigurable control of nonlinear optical devices through tunable spectral shaping with active pump modulation[49]. These findings, supported by theoretical analyses and numerical simulations across various platforms, underscore the potential of FWM-based transduction to deliver distinct efficiency and fidelity benefits under optimized operational regimes, positioning it as a highly attractive solution for scalable quantum information networks. Implementing FWM effectively, however, requires a physical platform capable of sustaining high intensities, providing tunable dispersion, and maintaining excellent optical coherence. Gas-filled hollow-core capillary fibers (HCFs) are particularly well-suited for this role[50,51].

The spectral phase of light pulses plays a fundamental role in quantum optics, as it directly governs the temporal and spectral structure of a photonic wavepacket. In quantum information science, where information is encoded not only in photon number but also in the coherent superposition of frequency and time modes, the spectral phase becomes a critical parameter for accurately describing and manipulating quantum states of light. The ability to reconstruct the complex spectral amplitude, including phase information, at the single-photon level provides key insight into the coherence properties of quantum states[52–54]. Spectral phase correlations between photons can also induce distinct quantum interference patterns, which are essential for applications such as boson sampling, quantum metrology, and quantum communication protocols[55,56]. Therefore, understanding and controlling spectral phase in frequency conversion processes is crucial for advancing high-performance and phase-coherent quantum transduction schemes.

Historically, gas-filled HCFs have been used for the development of high-performance light sources by enabling pulse compression[50,57,58]. In this work, we investigate spectral phase mapping in three distinct frequency conversion regimes: (1) input infrared (IR) to output ultraviolet (UV) frequencies within a gas-filled HCF system, (2) telecom-band (1550 nm) to UV (308 nm), and (3) telecom to visible (516 nm). These regimes are pre-selected to address spectral mismatches among various quantum platforms as examples. For instance, UV frequency comb generation for precision spectroscopy and optical clock platforms, which play a foundational role in quantum metrology and next-generation atomic clocks[59,60]. Trapped-ion qubits and Rydberg-atom systems often operate in the UV region[61–63], while nitrogen-vacancy (NV) centers in diamond and certain rare-earth-ion-doped crystals exhibit optical transitions near 516 nm[64–66]. Meanwhile, telecom-band photons are ideal for long-distance fiber-based quantum communication due to their low-loss transmission in standard optical fibers[67]. Therefore, realizing coherent and low-noise transduction between telecom and UV/visible domains is a critical step toward enabling heterogeneous quantum networks.

To benchmark the phase preservation across these regimes, we focus on degenerate FWM, as shown in the standard system schematic (Figure 1a). This simplified configuration involves only two states (Figure 1b), offering a simplified platform for analyzing phase evolution and optimizing nonlinear coupling dynamics. The gas-filled HCF environment allows precise control over key external parameters, such as gas pressure and temperature, to achieve phase matching and minimize dispersion, both of which are essential for maintaining high spectral phase coherence and thus high-fidelity wavefunction phase transfer.

Figure 1c conceptually illustrates different orders of nonlinear optical processes and their associated Hamiltonians. The FWM process studied here is governed by third-order nonlinearity ($\chi 3$), associated with a quartic Hamiltonian. This facilitates strong parametric coupling between intermediary field, signal field, and idler fields, enabling efficient frequency conversion with high phase fidelity.

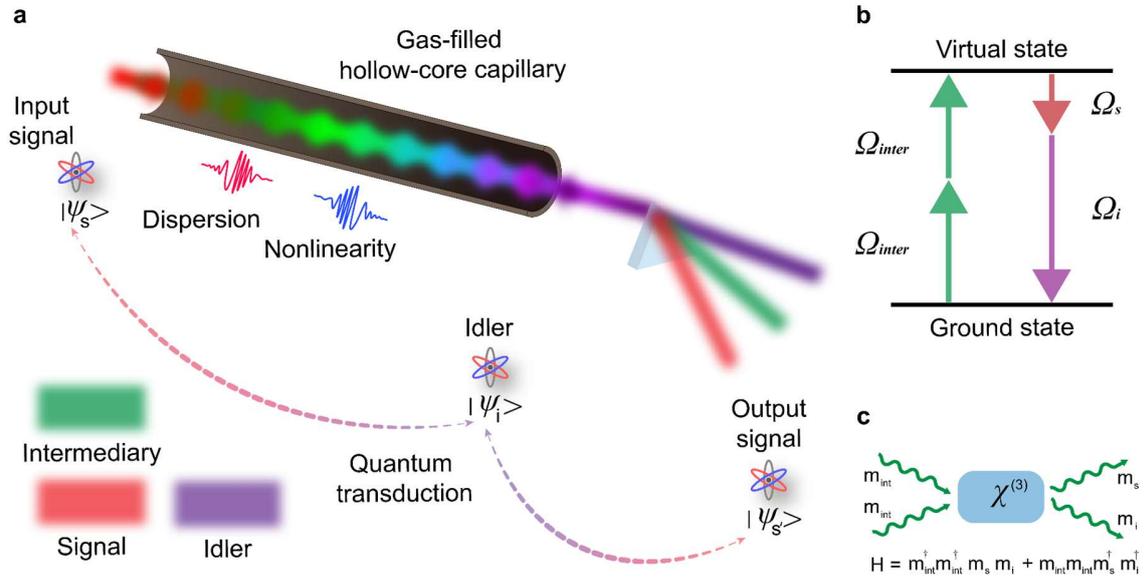

Figure 1. (a) Schematic representation of quantum transduction via four-wave mixing (FWM) in a gas-filled hollow-core waveguide. An intermediary laser interacts with an input beam through nonlinear optical processes, generating an output field. (b) The energy level diagram of the frequency mixing process, where two intermediary photons mediate the interaction between the input and output field phoçtons through a virtual state. (c) The diagrams and Hamiltonians of $\chi 3$ FWM process

## 2. Results

Here, we investigate coherent spectral phase transfer in three representative wavelength conversion regimes, each involving distinct combinations of input, intermediary, and output fields. These regimes are selected to reflect practically relevant scenarios in photonic quantum transduction where spectral mismatches commonly arise between quantum systems and communication channels.

Table 1. Selected nonlinear wavelength-conversion schemes and corresponding target applications.

| Regime | Wavelengths (Input → Intermediary → Output) | Applications |
|---|---|---|
| IR-to-UV | 1030 nm → 515 nm → 343 nm | Optical clocks, UV frequency combs (quantum metrology)[59,60] |
| Telecom-to-UV | 1550 nm → 515 nm → 308 nm | Trapped-ion & Rydberg systems (UV transitions for control/cooling)[61–63] |
| Telecom-to-Visible | 1550 nm → 775 nm → 516 nm | NV centers, rare-earth ions (quantum memories, repeaters)[64–66] |

Throughout our calculations, the input signal energy is kept constant at a quasi-single photon level ($10^{-12}$ μJ), while the intermediary pulse energy is varied between 4 μJ and 40 μJ. A comprehensive summary of the external physical parameters of the setup can be found in Table 1. At general ambient conditions (T = 293 K), phase matching is achieved when the fibers are filled with xenon gas. To ensure consistency and enable meaningful comparisons across our simulations, the input signal energy was kept constant throughout. This strategy also establishes a stable reference point for evaluating ablation effects and allows us to systematically assess conversion efficiency, which we define as the ratio between the output idler field energy and the input signal energy.

Table 2. Parameter settings in the FWM quantum transduction setup

| Parameter | Range/ Value |
| --- | --- |
| Temperature | 293 K |
| Fiber radius | 50 μm |
| Fiber length | 1 m |
| Gas medium | Xe |

To explore how different spectral phase structures affect the transduction fidelity and quantum coherence of photonic states, we focus on three representative phase profiles: linear, positive quadratic, and negative quadratic. These were chosen because they correspond to common physical scenarios (such as delay lines and dispersive media) and are readily implementable in practice. More importantly, each of these modulations alters the spectral-temporal structure of the wavefunction in distinct ways, thereby influencing key quantum properties like coherence, indistinguishability, and interference behavior. The table 3 below summarizes their physical interpretations and quantum mechanical implications.

Table 3. Physical interpretations and quantum-state implications of the selected spectral phase profiles.

| Phase modulation | Physical meaning | Quantum relevance |
| --- | --- | --- |
| Linear | Constant group delay (temporal shift) | Maintains spectral and temporal mode structure; quantum coherence and indistinguishability are preserved [69]. |
| Quadratic (Positive) | Positive chirp (temporal stretching) | Induces temporal-spectral entanglement; increases temporal duration and reduces overlap in two-photon interference[70]. |
| Quadratic (Negative) | Negative chirp (temporal compression) | Narrows temporal profile; increases spectral bandwidth; modifies Schmidt modes, and can influence entanglement generation [71]. |

## 2.1 Linear Phase Modulation

We first investigate the impact of linear phase modulation on the coherence of input transduction in the gas-filled hollow-core capillary, focusing on the IR-to-UV frequency conversion (case 1). Figures 2a–c present the input/output spectra and phases of the intermediary, signal, and idler, respectively. This regime serves as a baseline for evaluating phase preservation under near-linear conditions across a broad spectral span.

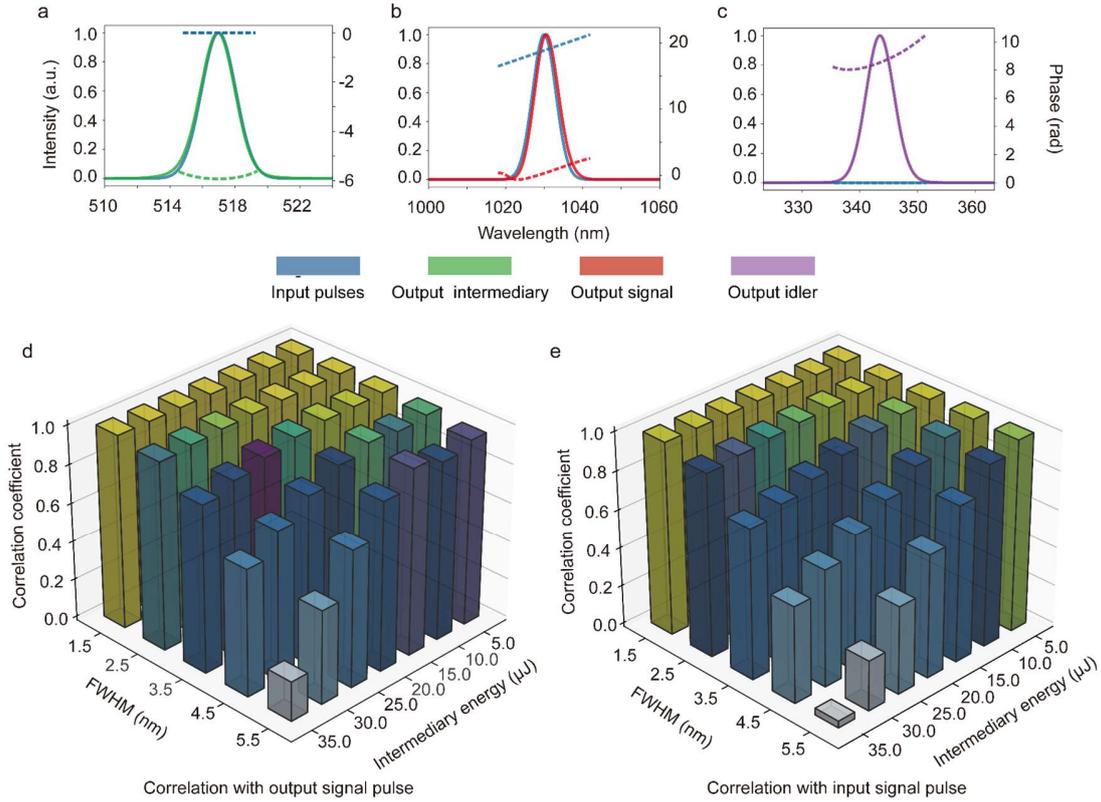

Figure 2. Linear-type phase mapping in a gas-filled hollow-core fiber (HCF): Examples of input/output spectrum and phase of (a) intermediary, (b) signal and (c) idler. (d) between the idler field and output signal pulses, and (e) between the idler output field and input signal pulse, as a function of varying input intermediary energy and the full-width at half-maximum (FWHM) of the intermediary pulse.

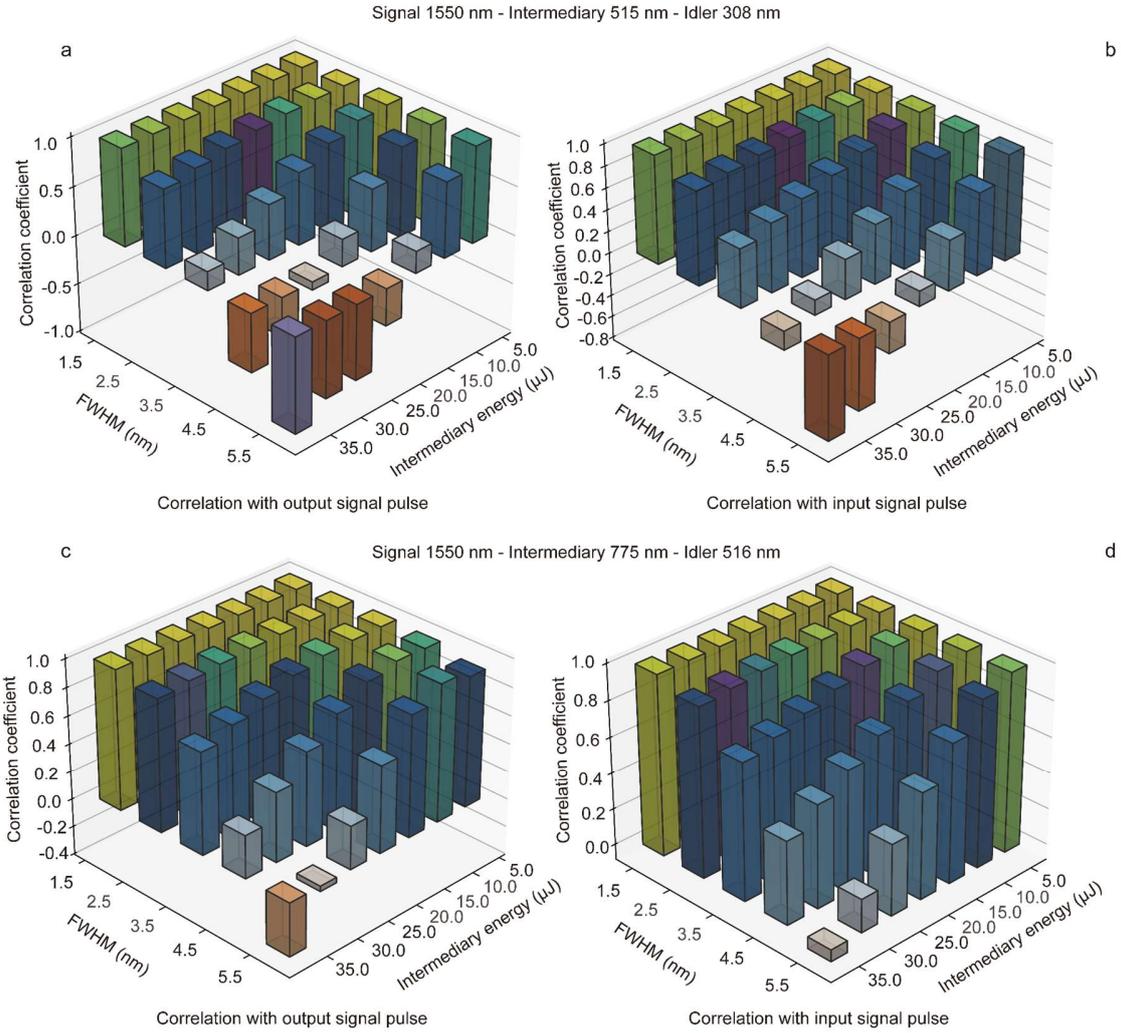

Figure 3. Spectral phase mapping correlations under (a) input 1550 nm, intermediary 515 nm and output field 303nm and (b) input 1550 nm, intermediary 775 nm and output field 516 nm: between the output output field and output input pulses, and between the output output field and input input pulse, as a function of varying input intermediary energy and the FWHM of the intermediary pulse.

To quantify the phase coherence and fidelity of the conversion process, we conducted a correlation analysis between the spectral phase of the input pulse and the generated output field pulses within their bandwidth. The correlation coefficient, which quantitatively evaluates the fidelity of phase transfer, spans from -1 to 1, with 1 representing perfect coherence and -1 indicating complete anti-correlation, reflecting a total loss of phase mapping. Figure 2d illustrates the correlation between the signal pulse and the output idler field pulse across varying intermediary energy and initial pulse bandwidths (FWHM). Even at relatively high intermediary energy and broader input bandwidths, the correlation remains above 0.95. This high correlation reflects not only classical waveform fidelity, but also implies that the underlying quantum wavefunction's spectral-temporal coherence is maintained throughout the conversion process. The quantum mechanical analog of these manipulations suggests a deep connection between spectral-temporal degrees of freedom and intrinsic photonic quantum properties[72]. As the FWHM of the input pulse increases, the coherence of the phase mapping declines notably. This behavior arises because the transmitted spectral phase is predominantly governed by the bandwidth of the intermediary, which is determined by the convolution of the input pulse spectrum with the intermediary spectrum itself. Similarly, increasing the intermediary energy leads to a sharp reduction in phase mapping coherence. This degradation is attributed to the onset of nonlinear effects, such as self-phase modulation (SPM) and cross-phase modulation (XPM), which induce spectral broadening of the intermediary and subsequently disrupt the fidelity of phase transfer during the conversion process.

Additionally, Figure 2e presents the correlation between the input pulse and the output field. The high correlation values (>0.99) at moderate intermediary energy (<10 μJ) and bandwidths (<1.5 nm) further confirm the effective transfer of phase information from the input to the output field frequency domain. As expected, at extremely high pulse energies and broad bandwidths, the correlation slightly degrades due to the onset of nonlinear distortions. Nevertheless, the overall high correlation values across a broad parameter space strongly support the preservation of spectral phase coherence through the parametric frequency mixing process.

Figure 3 presents the correlation analysis for two frequency conversion regimes: (a-b) 1550 nm to 308 nm and (c-d) 1550 nm to 516 nm. In both cases, we examine the correlation between the output idler field and the output signal (left) as well as the input signal (right). In the 1550–308 nm conversion (Figure 3a), high correlation (>0.99) is maintained at low intermediary energy and narrow FWHM, but degrades sharply as these parameters increase. This decline reflects the growing impact of nonlinear effects, particularly SPM and XPM, which distort the intermediary spectrum and disrupt spectral phase coherence. The correlation with the output signal is slightly more resilient than with the input signal, indicating partial preservation of internal symmetry despite overall phase degradation. For the 1550–516 nm case, the correlation remains consistently high (>0.95) across a broader range, with only moderate degradation at large FWHMs or high intermediary powers. This robustness arises from more favorable phase-matching conditions and reduced nonlinearity at longer wavelengths. Notably, both input-output field and output-output field correlations track similarly, confirming stable and symmetric phase transfer under realistic operating conditions.

## 2.2 Negative Quadratic Phase Modulation

As established in method section, when the initial phase modulation is an even-order function, the input and output fields acquire phases of the opposite sign. To further explore this behavior, we apply a negative quadratic phase modulation, corresponding to an up-chirped pulse, to the input and examine the resulting phase transfer characteristics.

Figure 4a–c shows an example of the spectral intensity and phase profiles of the input/output intermediary, signal, and idler fields under this condition, corresponding to the IR-to-UV conversion regime (case 1). The input intermediary (Figure 4a) remains nearly transform-limited with minimal phase modulation. In contrast, the signal pulse carries a clear negative quadratic phase, leading to an up-chirped profile (Figure 4b). The idler field (Figure 4c) develops a spectral phase that is approximately the mirror image of the signal's phase (positive quadratic phase), confirming the expected opposite-sign phase relationship dictated by the parametric mixing process.

To quantify the fidelity of the phase mapping, we analyze the correlation between the spectral phases of the output waves. Figure 4d presents the correlation between the idler field and output signal phases as a function of the intermediary energy and intermediary's bandwidth in the IR-to-UV regime (case 1). Similarly, Figure 4e shows the correlation between the idler field and the original input signal field. In both cases, the correlation coefficients exhibit significant degradation in regions of high intermediary energy and broad bandwidths, indicating sensitivity to nonlinear effects, inherent loss, and phase distortion under these conditions. At lower energy and moderate bandwidths, the correlation remains high, affirming effective phase transfer. However, as both parameters increase, the correlation declines sharply due to the onset of nonlinear phenomena and dispersion, which perturb the spectral phase structure during propagation.

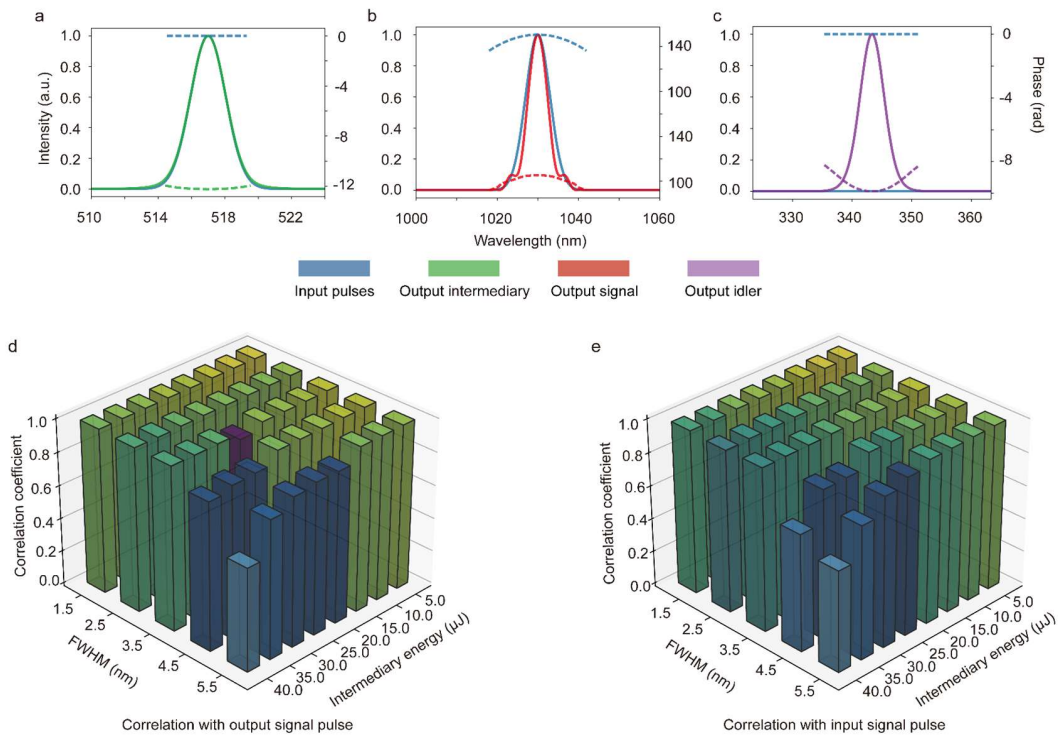

Figure 4. Negative quadratic (up-chirped) phase mapping in a gas-filled HCF: Examples of input/output spectrum and phase of (a) intermediary, (b) signal, and (c) idler. (d) between the idler field and output signal pulses, and (e) between the idler field and input signal pulse, as a function of varying input pulse energy and the FWHM of the intermediary pulse.

Figure 5 presents analogous correlation maps for case 2 (1550 nm to 308 nm) and case 3 (1550 nm to 516 nm). In Figure 5a and b, the 1550-to-308 nm conversion shows generally high correlation at low pulse energies, consistent with case 1. However, due to the larger frequency gap and higher nonlinearity required for UV generation, the degradation in both idler/output-signal field and idler/input-signal field correlations are more significant as the intermediary energy increases, particularly beyond 3. The correlation with the input drops below zero in the strong intermediary field regime, indicating loss of spectral phase fidelity due to intermediary distortion and excessive nonlinear broadening.

In contrast, Figure 5c-d, corresponding to the 1550-to-516 nm regime, exhibits a notably more stable performance. The correlation values for both output telecom–output visible field and input telecom–output telecom field phases remain above 0.95 across most of the tested parameter space, even up to 40 µJ. Only under the broadest bandwidth and strongest intermediary do we observe modest degradation, highlighting the relative robustness of the phase transfer in this regime. This improved stability is attributed to better phase-matching conditions and lower nonlinear thresholds at longer output field wavelengths.

Compared to the linear modulation, these findings show the sensitivity of phase mapping to both intermediary bandwidth and energy under negative quadratic modulation. While the phase inversion between the input and output field is clearly observed, maintaining high-fidelity transfer requires fine-tuning and optimization of operating parameters to mitigate nonlinear distortions.

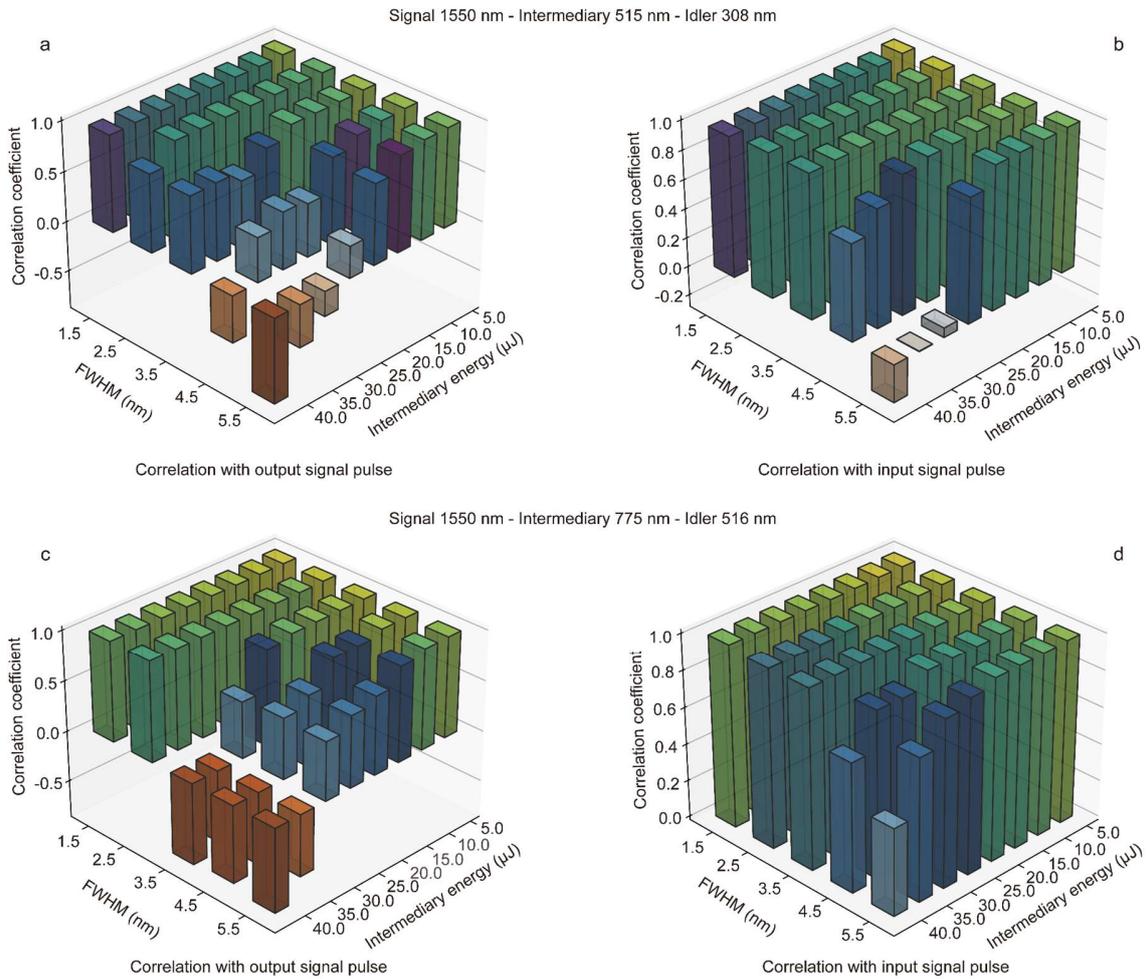

Figure 5 Spectral phase mapping correlations under (a-b) input 1550 nm, intermediary 515 nm and output field 303nm and (c-d) input 1550 nm, intermediary 775 nm and output field 516 nm: between the output output field and output input pulses, and between the output output field and input input pulse, as a function of varying input intermediary energy and the FWHM of the intermediary pulse.

**2.3 Positive Quadratic Phase Modulation**

These results present the critical role of input phase structure in determining conversion fidelity. To complete the picture, we proceed to investigate positive quadratic phase modulation, corresponding to a down-chirped input field, and its impact on phase mapping dynamics.

Figure 6a–c illustrates the spectral intensity and phase profiles for the input/output intermediary, input, and output field under this condition. The input pulse is modulated with a positive quadratic phase, producing a characteristic down-chirped profile (Figure 6b). As expected from the theoretical symmetry of parametric processes, the output field pulse (Figure 6c) acquires a spectral phase that mirrors the input's initial phase, exhibiting an opposite curvature. Meanwhile, the intermediary pulse (Figure 6a) remains close to transform-limited, providing a stable driving light for the frequency conversion.

To quantitatively assess the phase coherence of this process, we examine the correlation of spectral phases between the interacting pulses. Figure 6d presents the correlation between the output IR field and output UV field across varying intermediary energy and bandwidths. High correlation values are observed at lower energy (up to 25 µJ) and narrower bandwidths (~1.5 nm), indicating robust phase transfer. However, as both parameters increase, the correlation degrades. Figure 6e complements this analysis by showing the correlation between the output idler field and the input signal pulse. The trend is consistent: high correlations at lower operating energy conditions, with noticeable degradation at high pulse energies and broader intermediary bandwidths.

Figure 7 shows the corresponding correlation behavior for the telecom-to-UV (case 2) and telecom-to-visible (case 3) conversion regimes. As shown in Figures 7a and b, the 1550 nm to 308 nm conversion again exhibits high phase correlation at a narrow bandwidth (<1.5 nm) and a low input energy. However, the phase coherence rapidly deteriorates as the energy ratio increases beyond 4.0, with correlation values turning negative in both input-output field and input-output field comparisons. In contrast, Figure 7b demonstrates significantly improved robustness in the 1550 nm to 516 nm regime. Although a similar trend of decreasing correlation with increasing energy and bandwidth is observed, the overall phase coherence remains much higher. Even at energy ratios up to 4.0 and moderate FWHM, both correlation metrics stay above 0.94.

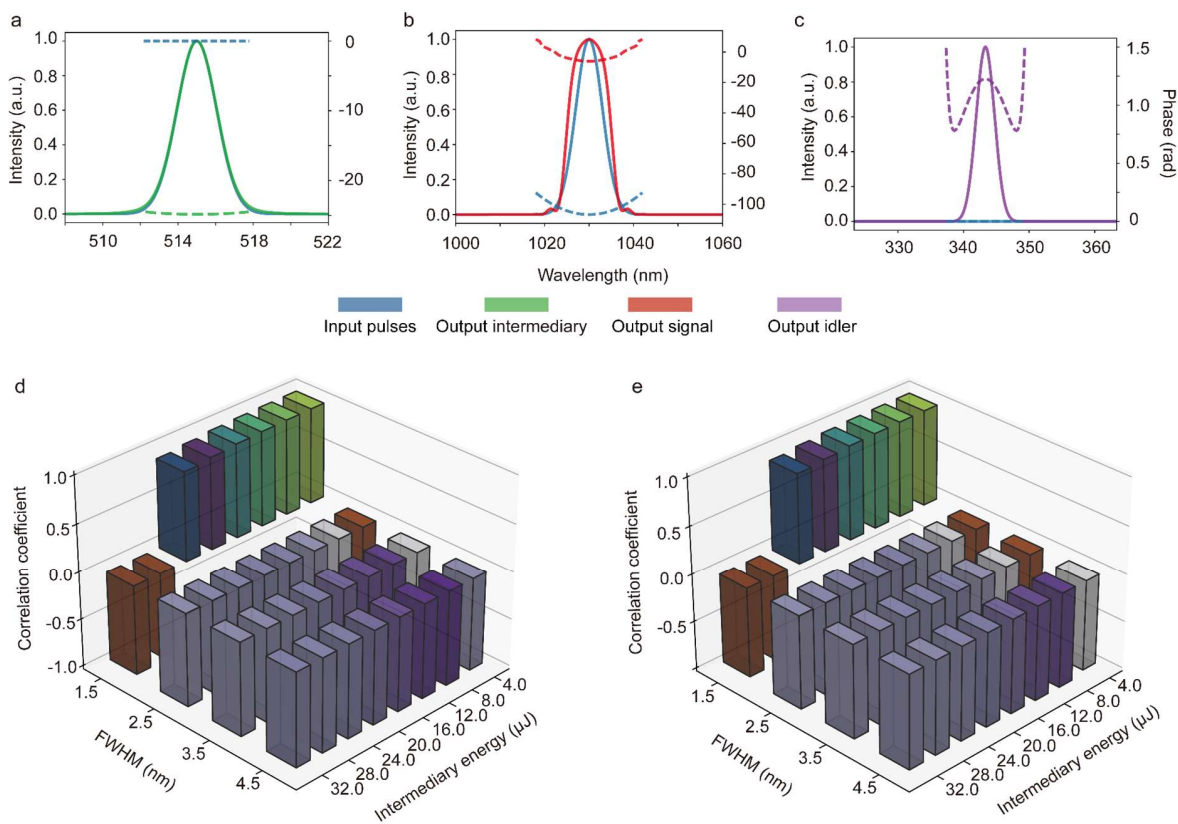

Figure 6. Positive quadratic (down-chirped) phase mapping in a gas-filled HCF: (a–c) Examples of input and output spectral distributions and phases mapping for (a) the intermediary, (b) the input, and (c) the output field. (d–f) Examples of pulse propagation dynamics along the HCF for (d) the intermediary, (e) the input, and (f) the output field. (g–h) Spectral phase mapping correlations: (g) between the output UV field and output IR pulses, and (h) between the output UV field and input IR pulse, as a function of varying input intermediary energy and the FWHM of the intermediary pulse.

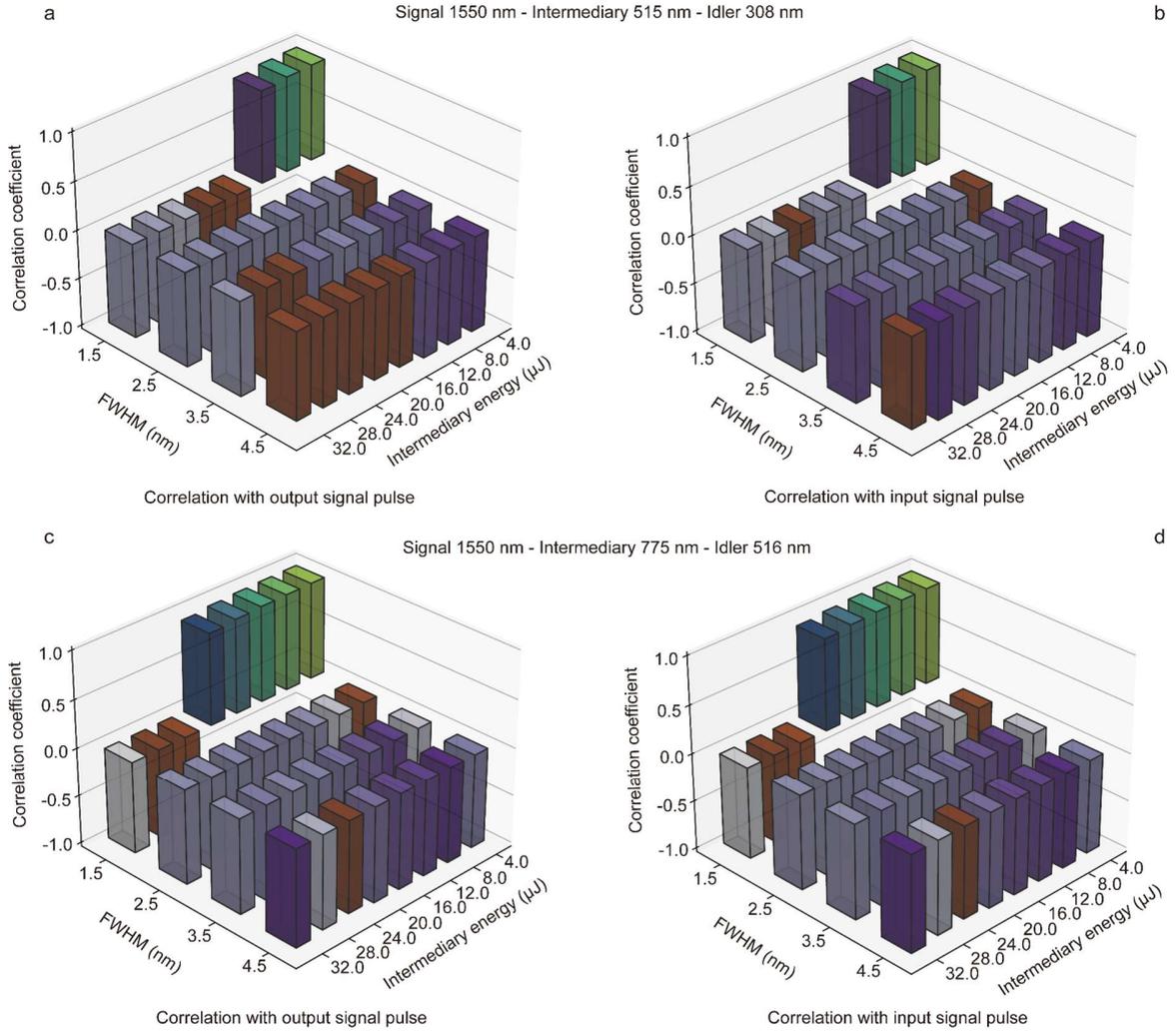

Figure 7 Spectral phase mapping correlations under (a-b) signal 1550 nm, intermediary 515 nm and idler field 303nm and (c-d) signal 1550 nm, intermediary 775 nm and idler field 516 nm: between the output idler field and output signal pulses, and between the output idler field and input signal pulse, as a function of varying input intermediary energy and the bandwidth.

It is also noteworthy that, in this case, the intermediary bandwidth is limited to the range of 1.5–4.5 nm, in contrast to the broader range explored in the previous cases. As shown in Figure 6d–e and Figure 7, the correlation coefficients drop rapidly when the intermediary bandwidth exceeds approximately 2 nm. This behavior arises because, in the case of positive quadratic modulation, the output field initially acquires an inverted (negative quadratic) phase structure relative to the input field. However, as the input intermediary energy increases, nonlinear effects such as SPM and XPM in the output field channel become increasingly significant. This is also related to the intermediary's monochromatic characteristic. These nonlinear interactions lead to the spectral broadening of the output field pulse, which inherently reshapes its phase profile toward a positive quadratic form. Consequently, the output spectral phase assumes a parabolic shape, and the corresponding pulse duration shortens due to the combined influence of nonlinear phase accumulation and group velocity dispersion. This phenomenon is less prominent in the previous negative-quadratic case, where either the phase modulation formats were matched or the nonlinear effects were less disruptive. In the positive quadratic modulation scenario, this interplay/compensation between nonlinear broadening and phase inversion becomes a dominant factor limiting the fidelity of phase transfer.

## 2.4 Conversion Efficiency Analysis

Following the investigation of phase mapping coherence under different modulation schemes, we now turn to evaluate the conversion efficiency across the same parameter space. Figures 8-10 compare the conversion efficiencies for three distinct phase modulation scenarios: (a) linear-type, (b) negative quadratic, and (c) positive quadratic, which show how the efficiency varies as a function of the intermediary's energy ratio and FWHM.

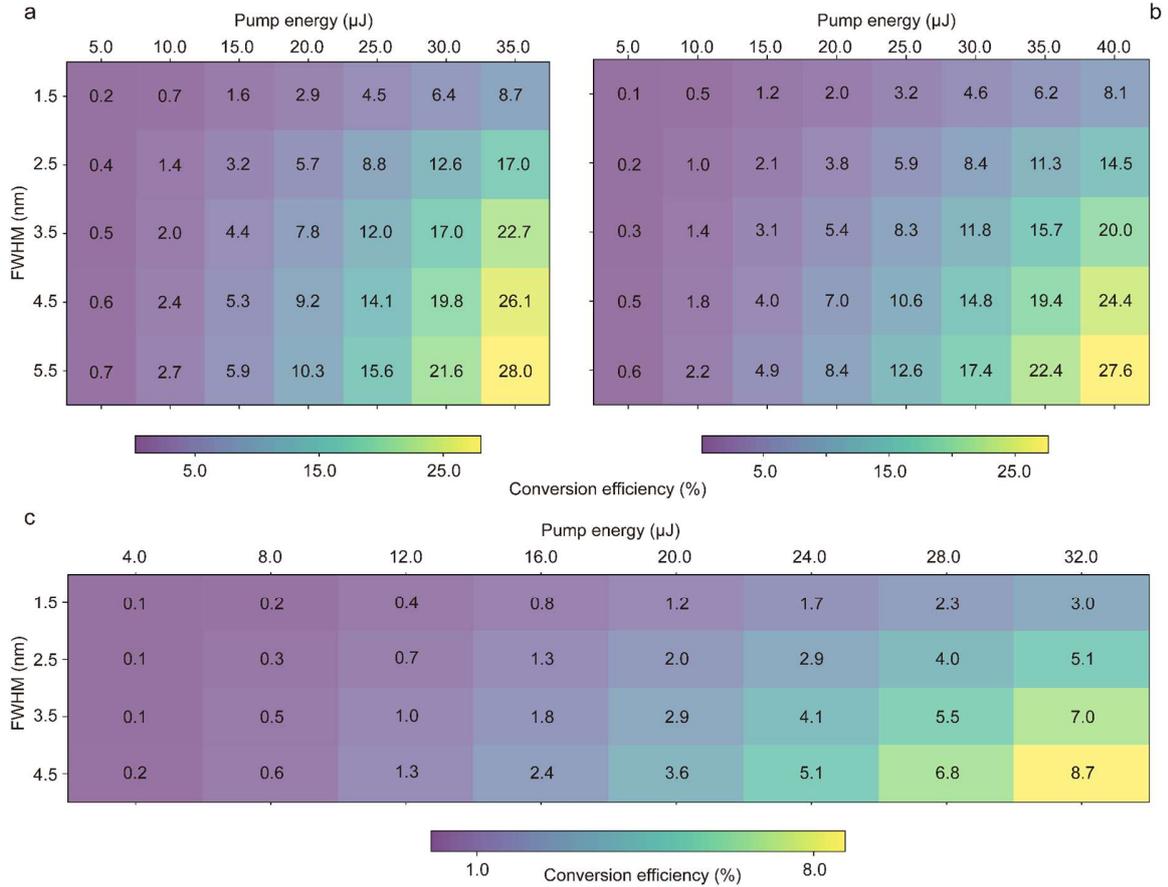

Figure 8. Conversion efficiency comparison for three different phase-matching cases: (a) linear-type, (b) negative quartic, and (c) positive quartic. The heatmaps illustrate the variation in conversion efficiency (%) as a function of the intermediary's energy and the FWHM. Higher conversion efficiencies will occur at a higher energy ratio with a large FWHM.

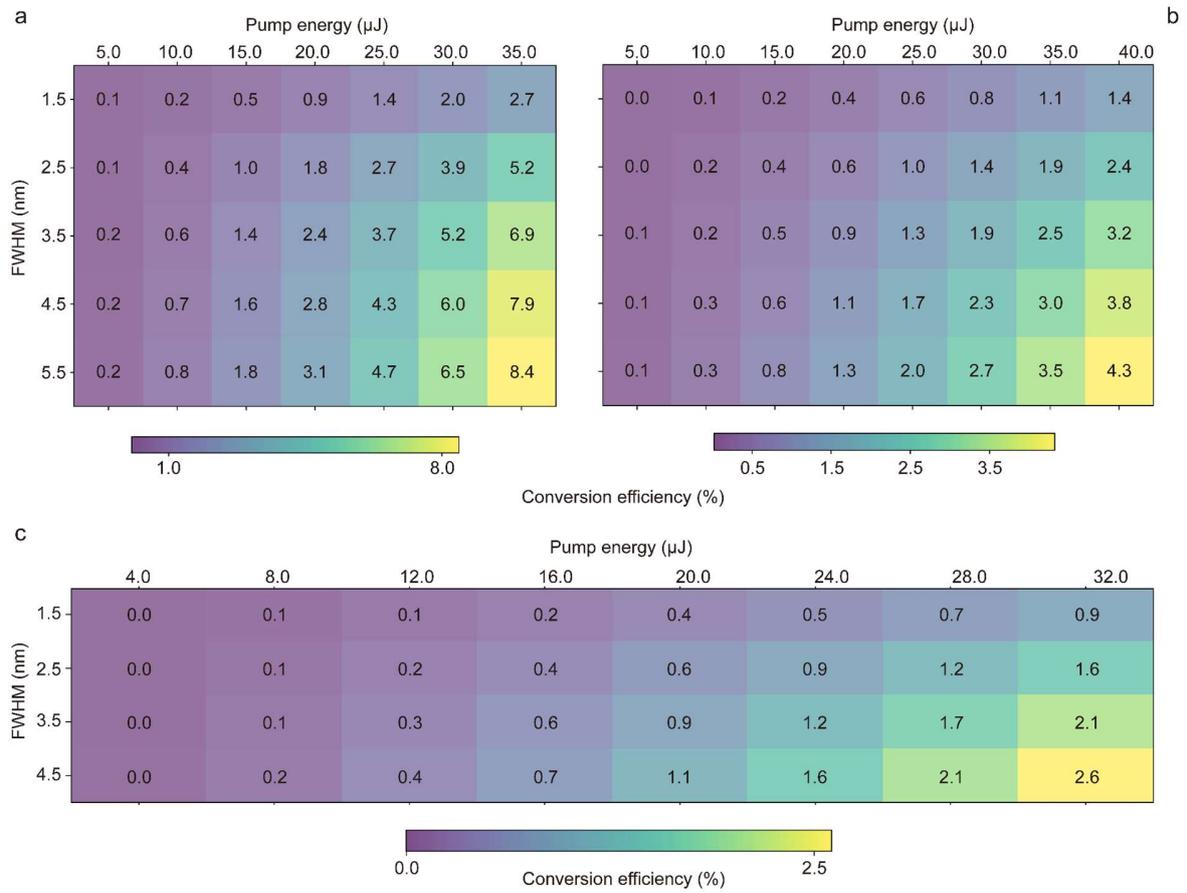

Figure 9 Conversion efficiency comparison for three different phase-matching cases: (a) linear-type, (b) negative quartic, and (c) positive quartic under input 1550 nm, intermediary 515 nm and output field 303nm case.

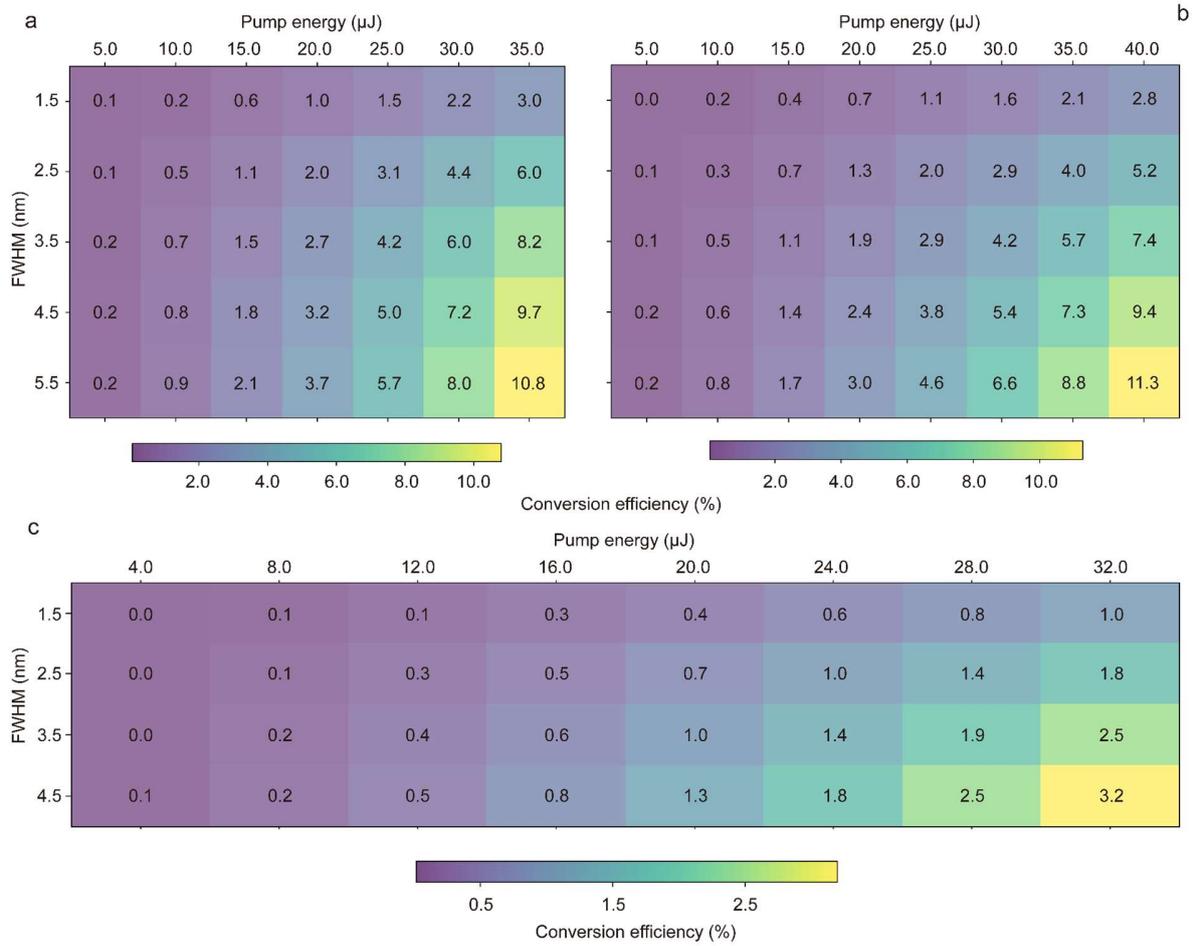

Figure 10 Conversion efficiency comparison for three different phase-matching cases: (a) linear-type, (b) negative quartic, and (c) positive quartic under input 1550 nm, intermediary 775 nm, and output field 516 nm case.

Consistent with our expectations, all three conversion regimes exhibit increasing conversion efficiency with higher intermediary energy and broader intermediary bandwidths. This trend reflects the enhanced nonlinear interaction strength and increased spectral overlap between the intermediary and input pulses under these conditions. In the IR-to-UV regime (case 1) at different intermediary energy level, the linear phase modulation case yields the highest conversion efficiency, reaching up to 28.0% (Figure 8a). Applying a negative quadratic phase reduces the efficiency to approximately 27.6% (Figure 8b), due to temporal broadening and reduced peak intensity. The positive quadratic phase case shows intermediate performance, achieving up to 8.4% efficiency (Figure 8c).

For the telecom-to-UV conversion (case 2), the overall conversion efficiencies are lower due to the increased phase mismatch and broader spectral separation. At different intermediary energy level, the maximum efficiency reaches 8.4% in the linear case, while negative and positive quadratic modulation reduce the values to 4.3% and 2.6%, respectively. In the telecom-to-visible regime (case 3), similar behavior is observed. The linear case achieves a moderate maximum efficiency of 10.8%, while the negative and positive quadratic cases reduce the performance to 11.3% and 3.2%, respectively.

However, this improvement in conversion efficiency comes with an important trade-off. As shown in our previous sections, higher intermediary energies and broader bandwidths — while beneficial for efficiency — lead to significant degradation in phase coherence. Specifically, in the positive quadratic modulation case, the strong nonlinear effects (SPM and XPM) not only broaden the output field spectrum but also distort the phase profile, undermining the fidelity of quantum information transfer. This highlights a crucial balance between achieving high conversion efficiency and maintaining phase coherence, which is essential for applications in coherent quantum state transduction.

Interestingly, the linear modulation case (Figures 8, 9, and 10a) offers the most favorable balance, delivering relatively high conversion efficiency while maintaining superior phase fidelity across a broad operational window. However, the impact of quadratic phase modulation differs across regimes. In the telecom-to-UV case, the negative quadratic modulation (Figure 9b) results in the lowest efficiency (~4%), likely due to temporal broadening that reduces the peak intensity and phase mismatch between the interacting waves. The positive quadratic modulation consistently leads to the lowest efficiency (2.6% and 3.2%, respectively), suggesting that the forward chirp introduces more detrimental temporal misalignment in these higher-frequency-separation conversions.

While increasing intermediary power and bandwidth can enhance nonlinear interaction and energy transfer, it also amplifies unwanted phase distortions, particularly under quadratic modulation. Therefore, achieving optimal performance requires a tailored parameter design, balancing conversion efficiency and phase coherence based on the priorities of the application, whether focused on energy throughput or on preserving quantum phase information during frequency conversion.

In addition to providing a comprehensive analysis of phase coherence and conversion efficiency, our study offers significant impacts for applications in quantum transduction. Although our present investigation is conducted in classical observation, the observed high-fidelity spectral phase mapping strongly indicates the potential for extending this approach to quantum states. Specifically, our findings demonstrate that the spectral phase, a critical parameter for preserving quantum coherence, can be reliably transferred across different frequency domains using a gas-filled HCF platform driven by FWM processes.

A particularly important insight from our work is the evidence suggesting that quantum information can be mapped effectively from the input field to the output frequency-converted field and that the high coherence between the newly generated output field and the input field can also be maintained. This broader applicability expands the potential for connecting quantum systems that operate in diverse spectral regions, such as linking telecommunications-band quantum memories with visible-to-UV light quantum processors or detectors. Traditional frequency conversion techniques often face severe challenges related to phase noise accumulation and limited efficiency[73–75]. In contrast, our results show that with appropriate control of the intermediary bandwidth and energy ratio, it is possible to achieve both high conversion efficiency and preserved phase fidelity, laying a practical foundation for coherent quantum frequency bridging.

Our comparison across linear, negative quadratic, and positive quadratic modulation schemes reveals critical trade-offs that must be considered in the design of future quantum transducers. For instance, while the quadratic modulation can deliver high conversion efficiency, it does so at the cost of significant phase distortion, especially at higher intermediary bandwidths and energies. This distortion arises from the compounded effects of SPM, XPM, and GVD, which reshape the output field's phase profile and diminish coherence. Conversely, the linear modulation format emerges as a more balanced configuration, sustaining both strong phase fidelity and reasonable efficiency across a broad operational window.

Furthermore, our analysis of the correlation coefficients provides insights into the underlying dynamics of spectral phase mapping. We observe that increasing the intermediary bandwidth beyond ~2 nm triggers a transition in the output field's phase structure, driven by nonlinear spectral broadening. In the positive quadratic case, this leads to a phase curvature inversion followed by restoration to a positive parabolic shape, accompanied by pulse compression. Such behavior, while challenging for maintaining perfect coherence, could be exploited in quantum transduction protocols that intentionally harness nonlinear reshaping effects for tailored quantum state preparation[76–78].

Moreover, a deeper understanding of these nonlinear-induced phase transformations could inform the development of quantum error corrections[79–81], where anticipated phase distortions are mitigated or actively compensated. By integrating appropriate error correction protocols that account for systematic phase shifts and spectral broadening, it may be possible to enhance the robustness of quantum information transfer, even in regimes where nonlinear effects are non-negligible. This not only validates the versatility of the gas-filled HCF platform for coherent frequency conversion but also charts a clear pathway for optimizing performance in future quantum applications across different platforms. By understanding and managing the balance between nonlinear enhancement and phase preservation, this approach holds strong promise for enabling robust, low-noise quantum networks[82–84]. As future work, experimental efforts focused on direct quantum state measurements and entanglement preservation will be critical to fully unlock the potential demonstrated here.

## 3. Discussion

This study has demonstrated the potential for implementing quantum transduction through coherent spectral phase mapping in a gas-filled hollow-core fiber platform. By systematically exploring three typical frequency conversion regimes, IR-to-UV, telecom-to-UV, and telecom-to-visible, we have shown that it is possible to transfer spectral phase information with high fidelity from the signal field to the generated idler field. Our analysis confirms not only the preservation of spectral phase coherence during the conversion process but also strong phase correlations between the input and output fields, highlighting the capability of our approach to support coherent quantum information transfer across disparate spectral domains. These findings go beyond conventional demonstrations of frequency conversion, offering new insights into the physical mechanisms necessary for phase-preserving quantum transduction. The comparative evaluation across the three regimes further reveals the distinct roles of spectral separation, nonlinear dynamics, and modulation format in determining the trade-offs between conversion efficiency and phase fidelity. With further experimental validation, particularly focusing on quantum state preservation and entanglement mapping, this platform holds strong promise for enabling practical, low-noise, and spectrally versatile quantum interfaces. Our results lay foundational groundwork for integrating heterogeneous quantum systems and advancing the development of coherent, scalable quantum networks.

## 4. Methods

**4.1 Theoretical Framework for Phase-Coherent Photonic Frequency Conversion**

Consider intermediary, signal, and idler fields in an FMW process with the annihilation operators $(\hat{a}_{int})$, $(\hat{a}_s)$, and $(\hat{a}_i)$, respectively. The Hamiltonian for this process can be written as [38,40,68]:

$$\hat{H}_{\text{QFWM}} \propto \hbar\kappa \left( \hat{a}_{int}^\dagger \hat{a}_{int}^\dagger \hat{a}_s \hat{a}_i + \text{H.c.} \right) \tag{1}$$

where $\kappa$ refers to the nonlinear coupling strength. Assuming the intermediary field light is a classical field with high intensity:

$$\hat{a}_{int} \rightarrow A_{int} \exp(i\phi_{int}) \tag{2}$$

The Hamiltonian simplifies to:

$$\hat{H}_{\text{QFWM}} = \hbar\kappa A_{int}^2 \left( \hat{a}_s \hat{a}_i + \hat{a}_s^\dagger \hat{a}_i^\dagger \right) \tag{3}$$

## 4.2 Phase Mapping Relationship

Using the Heisenberg equation, for the signal and idler fields, we have:

$$\frac{d\hat{a}_s}{dt} = \frac{i}{\hbar}\left[\hat{H}_{\text{QFWM}}, \hat{a}_s\right] \tag{4}$$

$$\frac{d\hat{a}_i}{dt} = \frac{i}{\hbar}\left[\hat{H}_{\text{QFWM}}, \hat{a}_i\right] \tag{5}$$

Transforming the time coordinate into the spatial coordinate, we establish the evolution relationship between the signal and idler fields:

$$\frac{d\hat{a}_s}{dz} = -i\gamma A_{int}^2 \hat{a}_i^\dagger \tag{6}$$

$$\frac{d\hat{a}_i}{dz} = -i\gamma A_{int}^2 \hat{a}_s^\dagger \tag{7}$$

where $\gamma$ is the transformed nonlinear coupling strength. With initial conditions $\hat{a}_s(0)$ and $\hat{a}_i(0)$, the solution is:

$$\hat{a}_s(z) = \cosh(gz)\hat{a}_s(0) + \sinh(gz)\hat{a}_i^\dagger(0) \tag{8}$$

$$\hat{a}_i(z) = \cosh(gz)\hat{a}_i(0) + \sinh(gz)\hat{a}_s^\dagger(0) \tag{9}$$

where the gain coefficient $g = \gamma A_{int}^2$. In order to determine the phase relationship between the signal and idler fields in the frequency domain, we need to start with the frequency representation of the annihilation operators for the signal and idler fields. Consider the initial fields at the input of the waveguide, the annihilation operators for the signal and idler fields can be expressed in terms of their spectral amplitudes and phases as follows:

$$\hat{a}_{s0}(\omega) = |\hat{a}_{s0}(\omega)|e^{i\phi_{s0}(\omega)} \tag{10}$$

$$\hat{a}_{i0}(\omega) = |\hat{a}_{i0}(\omega)|e^{i\phi_{i0}(\omega)} \tag{11}$$

Through the degenerate FWM process in the waveguide, the fields evolve along the propagation direction Z. The annihilation operators at the output end $(z = L)$ are given by:

$$\hat{a}_s(\omega, L) \approx \cosh(gL)\hat{a}_{s0}(\omega) + \sinh(gL)\hat{a}_{i0}^\dagger(\omega) \tag{12}$$

$$\hat{a}_i^\dagger(\omega, L) \approx \sinh(gL)\hat{a}_{s0}(\omega) + \cosh(gL)\hat{a}_{i0}^\dagger(\omega) \tag{13}$$

For simplicity, we neglect the absorption loss in the waveguide. By substituting the initial annihilation operator expressions into the above equations, we get:

$$\hat{a}_s(\omega, L) \approx \cosh(gL)|\hat{a}_{s0}(\omega)|e^{i\phi_{s0}(\omega)} + \sinh(gL)|\hat{a}_{i0}(\omega)|e^{-i\phi_{i0}(\omega)} \tag{14}$$

$$\hat{a}_i^\dagger(\omega, L) \approx \sinh(gL)|\hat{a}_{s0}(\omega)|e^{-i\phi_{s0}(\omega)} + \cosh(gL)|\hat{a}_{i0}(\omega)|e^{i\phi_{i0}(\omega)} \tag{15}$$

Herein, we find that the phases $\phi_{s0}$ and $\phi_{i0}$ of the input annihilation operators directly influence the phase of the output operators:

$$\phi_s(\omega, L) \approx \arg\left(\cosh(gL)e^{i\phi_{s0}(\omega)} + \sinh(gL)e^{-i\phi_{i0}(\omega)}\right) \tag{16}$$

$$\phi_i(\omega, L) \approx \arg\left(\sinh(gL)e^{-i\phi_{s0}(\omega)} + \cosh(gL)e^{i\phi_{i0}(\omega)}\right) \tag{17}$$

Thus, we can see that the phase of the idler field $(\phi_i(\omega, L))$ is influenced by both the phase of the initial signal $(\phi_{s0}(\omega))$ and the phase of the initial idler field $(\phi_{i0}(\omega))$, where the initial idler field term is 0.

**4.3 Odd and Even Phase Mapping Relationship**

A signal at $\omega_s + \omega$ generate idler $\omega_i - \omega$. A signal component at $\omega_s - \omega$ generates idler $\omega_i + \omega$:

$$\phi_i(\omega_i - \omega) = 2\phi_{int} - \phi_s(\omega_s + \omega) \tag{18}$$

Assuming $\delta = -\omega$, and expending $\phi_s$ at $\omega_s$, we have:

$$\phi_s(\omega_s - \delta) = \phi_s(\omega_s) - \delta\phi_s^{'}(\omega_s) + \frac{\delta^2}{2}\phi_s^{''} - \frac{\delta^3}{6}\phi_s^{'''}... \tag{19}$$

So the idler phase could be calculated by:

$$\phi_i(\omega_i + \delta) = 2\phi_{int} - \phi_s(\omega_s) + \delta\phi_s^{'}(\omega_s) - \frac{\delta^2}{2}\phi_s^{''} + \frac{\delta^3}{6}\phi_s^{'''}... \tag{20}$$

If the phase terms are odd functions, the signal and idler fields exhibit phases with the same sign. Conversely, if the phase terms are even functions, then the signal and idler field operators display phases of opposite signs.

**Acknowledgment.** The author thanks the support from UCLA and SLAC National Accelerator Laboratory, the U.S. Department of Energy (DOE), the Office of Science, Office of Basic Energy Sciences.

**Funding.** The U.S. Department of Energy (DOE), the Office of Science, Office of Basic Energy Sciences under Contract No. DE-AC02-76SF00515, No. DE-SC0022559, No. DE-FOA-0002859, the National Science Foundation under Contract No. 2231334.

**Declarations.** The authors declare no competing interests.

**Data availability.** Data underlying the results presented in this paper are not publicly available at this time but may be obtained from the authors upon reasonable request.

**Contributions.** The study was conceived and designed by H.Z., S.C. Data analysis and result interpretation were conducted by H.Z., Y.X., W.C., L.S., R.W.B. S.C. supervised the analysis, simulations, and theoretical framework of the study. All authors contributed to the writing and revision of the manuscript.

22